# A New Flexible Modified Impedance Network Converter

Shirin Besati[1,3], Somasundaram Essakiappan[2], and Madhav Manjrekar[1,3]
[1]Department of Electrical and Computer Engineering, University of North Carolina at Charlotte, USA
[2]QM Power Inc, Charlotte, USA
[3]Energy Production and Infrastructure Center, University of North Carolina at Charlotte, USA
sbesati@uncc.edu, somasundaram@qmpower.com, mmanjrek@uncc.edu

*Abstract*— One of the popular impedance-network converters are Y-source converters which along with their essential characteristics such as reducing the size of converter components, single-stage power transferring, fault tolerance, and wide voltage gain capabilities there are also some drawbacks that one of the most widespread is high leakage inductances which affect performance negatively. This paper introduces a new configuration based on coupled inductors as a power electronic converter in three case studies to verify the high reliability of the proposed converter by using just a simple controller that is substantial for recycling brake energy in the propulsion system of Electric Vehicles and grid-following inverters. This topology with a straightforward structure, computation, and wide voltage gain, provides a proper connection between the components of its network that obtains appropriate paths for leakage energy and likewise helps soft-switching in some conditions. Additionally, the performance of other previously related constructions is compared with the suggested topology. Simulations based on MATLAB/ SIMULINK have been carried out, and correspondingly experiment results have been also presented to substantiate the theoretical outcomes.

*Keywords*— impedance-network converters, coupled inductors, Y-source converters, Electric Vehicles, grid-following inverters

## I. INTRODUCTION

Considering the growth in the usage and application of power electronic converters in Electric Vehicles (EV), Robotics, Energy Storage, Smart Grid Technology, islanded systems, and so many other applications in the industry, we are attracted to optimizing the performance of the power electronic converters [1-6]. To the best of our knowledge, the classical impedance-network converter (ZSC) was first demonstrated by F.Z. Peng in 2003 to overcome the limitations of conventional inverters: Voltage Source Converter (VSC) and Current Source Converter (ISC) [7]. Nowadays, a huge number of highly capable power converter topologies for power quality improvement applications have been illustrated. There are unique features for impedance network topologies that are not seen in regular converters. The impedance network with a buck/boost (BF) capability, provides single-stage DC-AC, AC-DC, DC-DC, and AC-AC conversions. In ZSI both power switches of a leg can be turned on simultaneously, therefore, dead time is not a concern anymore. This significantly improves reliability and reduces output waveform distortion. Consequently, the operation period of these converters, unlike conventional structures, consists of two parts; active state (NST) and shoot-through state (ST) [7, 8]. Along with their prominent specifications, a classical ZSC also has its weaknesses such as (1) a basic ZSC uses two capacitors in its network and causes inrush current at startup; (2) they regulate gain factor only by adjusting the shoot-through duty ratio, this is while numerous different impedance-network converters have been purposefully proposed to solve these disadvantages [9-13]. In this paper, two previous ZSCs which are improved Y-source and modified Y-source inverters (YZSI), are briefly introduced and then they will be compared with the proposed topology. [14, 15]. In sum, the first topology in Fig. 1 has been proposed with greater flexibility of the number of coupled inductors in boost facto to improve the gain voltage (it is considered BF = 2). So, the boost factor has been changed from equation (1) pertains to a Classical Y-source converter (classical YZSI) to equation (2).

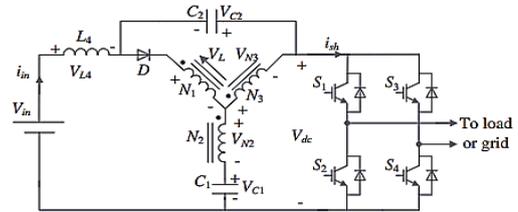

Fig. 1 Improved Y-source inverter.

$$B = \frac{1}{1-K.d} \quad (1)$$

$$K = \frac{(N_3+N_1)}{N_3-N_2}$$

$$N_{12} = \frac{N_1}{N_2}$$

$$N_{13} = \frac{N_1}{N_3}$$

$$B = \frac{V_{dc}}{V_{in}} = \frac{1}{1-(K+1)d} \quad (2)$$

Following upgrading this converter, Modified YZSI with the same boost factor, and by adding two capacitors, one diode, and another inductor to the impedance network for creating Additional pathways in order to diminish leakage energy, therefore, it has been able to reduce power losses. This topology also is able to eliminate the high voltage and current values at start-up to enhance efficiency and produce high reliability for switches and components within the network [15]. Likewise, it has complex calculations to provide a boost factor, and also, increasing the number of network components of this modified converter rises the cost. Fig. 2 shows the structure of Modified YZSI and the equations of it is presented in equation (3).
The main contributions of this paper are: (i) we designed a new ZSC converter with its distinctive features and compared it with the previous ones, (ii) we analytically computed the output/input voltage ratio (BF), (iii) we simulated the proposed converter's performance in SIMULINK/ MATLAB, (iv) we derived experimental results and compared with the simulation results.

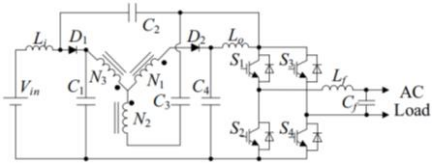

Fig. 2 Modified-Y-source inverter.

$K = \frac{(N3+N1)}{N3-N2}$

$N_{12} = \frac{N1}{N2}$

$N_{13} = \frac{N1}{N3}$

$B = \frac{Vdc}{Vin} = \frac{1}{1-(K+1)d}$ (3)

## II. PROPOSED TOPOLOGY

### 1. A. CIRCUIT INTRODUCTION

The proposed topology is indicated in Fig. 3, the number of impedance network components is equal to the first introduced structure and less than the second one. Here, a coupled inductor Y-source and a simple inductor are used for designing the impedance construction. The foremost objective for the new YZSI is expanding the boost factor by simple calculation, and improving efficiency along with saving continuous input current and the other properties of an impedance network converter. As mentioned before, the unique feature of this topology compared to the two previous converters is due to the particular design of the coupled inductors, it has been able to maintain the same number of improved Y-Source components, create proper pathways for the leakage energy and consequently increase efficiency. In this paper, the gain voltage has been considered as 5. Another exclusivity of the topology is the inductor presence in two branches connected to the switching bridge that contributes to the soft switching during inductive loads.

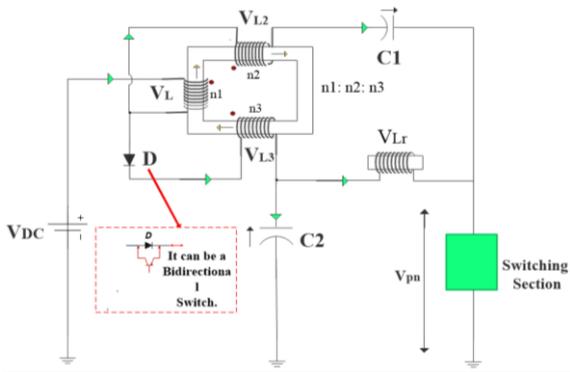

Fig. 3 Configuration of the proposed topology.

### 2. B. COMPUTATION PRINCIPLE

Fig. 4 shows the proposed topology equivalent circuit in the NST functional mode. By using the volt-second balance principle in equations (4), and (5), as a result, equations (6) and (7) are the voltages of $L_r$ and $L_1$ in terms of capacitor voltage and DC input voltage. Anew, the calculation is applied according to the equivalent circuit of the desired impedance network in ST mode in Fig.5 to obtain voltages of the network inductors in this term. In the manner of the routes in the ST equivalent circuit and equations (8), and (9), the network inductors are charging by discharging voltages of the capacitors as well as the DC input voltage, while diode 1 ($D_1$) is in reverse.

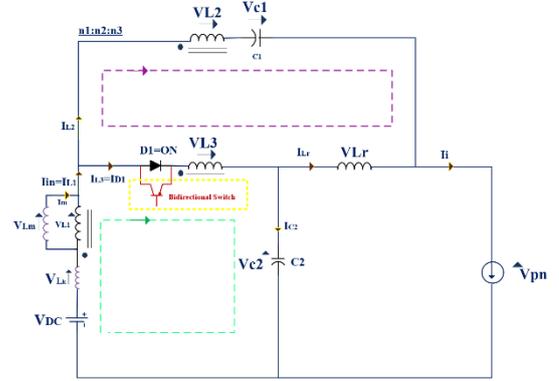

Fig. 4 Equivalent circuit of the proposed topology in NST mode.

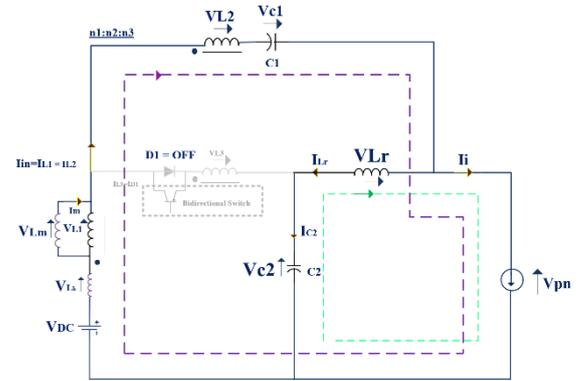

Fig. 5 Equivalent Circuit of the proposed converter in ST mode.

$\int_0^{Ts} VL1.dt = 0 \rightarrow \int_0^{Tst} VL.dt + \int_0^{Tnst} VL.dt = 0$ (4)

$\int_0^{Ts} VLr.dt = 0 \rightarrow \int_0^{Tst} VLr.dt + \int_0^{Tnst} VLr.dt = 0$ (5)

$K = \frac{n2}{n1}$, $P = \frac{n3}{n1}$

$V_{L1}^{NST} = \frac{1}{1+K} V_{DC} - \frac{1}{1+K} V_{C2}$ (6)

$V_{Lr}^{NST} = \frac{P-K}{1+K} V_{DC} - \frac{P-K}{1+K} V_{C2} - V_{C1}$ (7)

The voltages across C1 and C2 are obtained as follows,

$V_{C1} = \frac{d(1+K)}{(1-d).(1+P)-d.(1+K)} V_{DC}$ (8)

$V_{C2} = \frac{(1-d).(1+P)}{(1-d).(1+P)-d.(1+K)} V_{DC}$ (9)

Finally, DC link voltage and BF by replacing the voltage of the capacitor in the calculated equation of in NST mode, are obtained as the following equations.

$V_{pn} = V_{C1} + \frac{1+P}{1+K} V_{C2} - \frac{P-K}{1+K} V_{DC}$ (10)

$V_{pn} = \frac{(1+P)}{(1-d).(1+P)-d.(1+K)} V_{DC}$ (11)

$B = \frac{(1+P)}{(1-d).(1+P)-d.(1+K)}$ (12)

As well, the AC output voltage with M modulation coefficient for the proposed topology is equal to equation

(13). As can be seen from BF equation (12), the turn rations of three coupled inductors are present in the voltage gain equation. Concisely, Table 1 compares the proposed structure and previous Y-Source converters.

$$V_{AC} = \frac{M}{2} \cdot \frac{(1+P)}{(1-d).(1+P)-d.(1+K)} V_{DC} \qquad (13)$$

3. TABLE 1. COMPARISON BETWEEN THE PROPOSED STRUCTURE AND PREVIOUS Y-SOURCE CONVERTERS.

| | *Proposed topology* | *Improved YZSI* | *Modified YZSI* |
|---|---|---|---|
| *Number of capacitors* | 2 | 2 | 4 |
| *Number of inductors* | _ Y-Source winding _one inductor | _ Y-Source winding _one inductor | _ Y-Source winding _two inductors |
| *Number of diodes in impedance network* | One diode | One diode | 2 diodes |
| *Continuous input current* | Yes | Yes | Yes |
| *Startup inrush current* | Yes | No | - |
| *Soft switching* | Yes, it helps to soft switching in induction loads. | No | No |
| *Common grounding* | Yes | Yes | Yes |
| *Efficiency* | Well | Low | well |

## III. PERFORMANCE EVALUATION

In this section, the proposed converter is simulated in SIMULINK/ MATLAB software for three different loads that are inductive motor, inductive generator, and resistive load correspondingly by designing a simple unipolar pulse-width modulation (PWM). Three case studies have been presented to verify the high reliability of this converter by using just a simple controller, and additionally, to validate the theoretical results. Finally, the proposed topology with a 20 kHz switching frequency has been prepared as a lab sample for DC-DC conversion. The simulation characteristics for three cases are provided in TABLE 2.

TABLE 2. SIMULATION SPECIFICATIONS.

| Parameters of the proposed impedance network inverter | Amount | 3-phase induction machine | Amounts |
|---|---|---|---|
| $C_1$, $C_2$ | 100 μF | Voltage | 400 Volts |
| Y-source inductor | $n_1:n_2:n_3= 1:3.4:1$ $L_m= 370$ μH $L_{leakage}= 0.15$ μH | Power | 3.4 kW |
| Output inductor | 1 mH | Frequency | 50 Hz |
| $d_{st}$ | 25% | Speed | 1440 rpm |
| Input voltage | 80 V | Stator resistor | 2.125 Ω |
| Switching frequency | 20 kHz | Rotor resistor | 2.05 Ω |
| | | Stator inductance | 2 mH |
| | | Rotor inductance | 2 mH |
| | | Magnetic inductance | 6.4 mH |
| | | J | 0.015 kgm |

### 4. A. CASE 1: INDUCTION MOTOR

Fig. 6 presents the simulation results for the first case study with an induction motor as an inductive load and the input DC voltage is considered 80 V as Fig. 6(a). According to Figs. 6(b, c), the waveform curves are illustrating a continuous input current and current through the diode, severally where inrush currents at the beginning exist due to the inductors and the capacitors within the network. Fig. 6(d) shows the link voltage of the inverter with BF= 5, and by looking at the switch current in Fig. 6(e) Soft-switching performance has been proved. Moreover, Fig. 6(f) is the output results belonging to the motor with a positive torque.

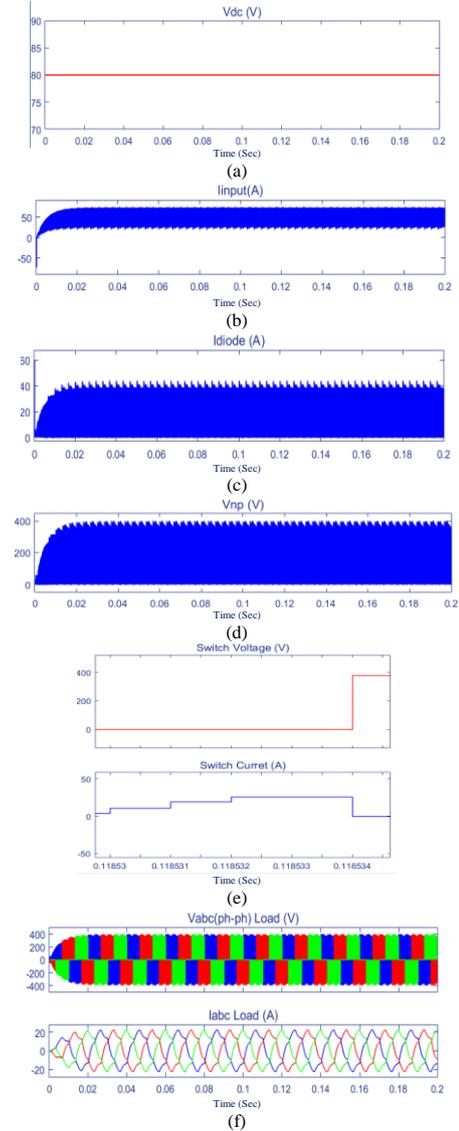

Fig. 6 Simulation results for induction motor.

### 5. B. CASE 2: INDUCTION GENERATOR

Case 2 is produced in an opposite direction (bidirectional path) for charging a battery by an induction generator with quantities exactly equal to case 1 and furthermore, the simulation results in this case are the same as the first episode. This similarity confirms the appropriate behavior and stability of the topology in different conditions and for different loads with only a simple control system. This aspect is important for recycling brake energy in EVs and grid-following inverters. The outcomes of this case study are presented in Fig. 7.

TABLE 3. PARAMETERS OF THE EXPERIMENTAL SAMPLE.

| *Parameters* | *Amount* |
|---|---|
| Input voltage | 20 Volts |
| Output voltage | 100 Volts |
| $L_r$ | 330 mH |
| Turn ration of Y-Source | n1:n2:n3= 1:2:2 |
| $C_1$ | 220 μF |
| $C_2$ | 680 μF |
| $R_{Load}$ | 245 Ω |
| $f_{switching}$ | 20 kHz |
| $d_{st}$ | 40% |

## 6. C. CASE 3: RESISTIVE LOAD AND LABORATORY RESULTS (DC-DC CONVERTER)

Due to the specification of impedance network converters that can be used in all power conversion modes, the proposed impedance network has been designed for a DC-DC converter with a resistive load. TABLE 3 and Fig. 8 determine the component quantities and experimental results in case study 3. Fig. 8(a) shows the input DC voltage is considered 20 volts, and the objective is to have a DC output of 100 volts which incomes the boost factor is 5, as deemed previously. Fig. 8(b) shows the continuous input current, and the other simulation results of the components have been presented in Fig. 8(c, d). As it can be seen from Fig. 8(e) a monolithic voltage waveform goes through $C_2$ because currents through both branches connected to this capacitor first pass through the inductors and then enter into the $C_2$. Correspondingly, The converter's efficiency is 98% in the results of the simulations and reality.

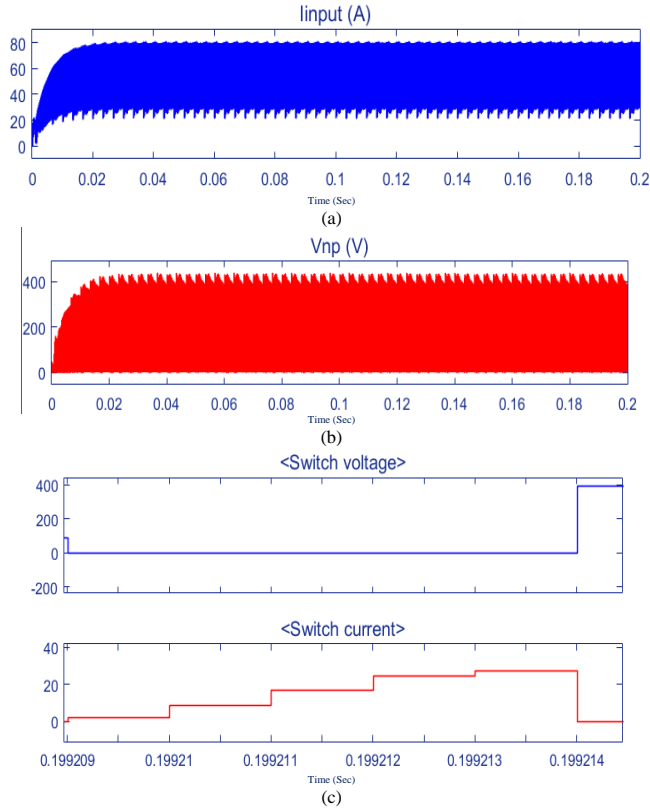

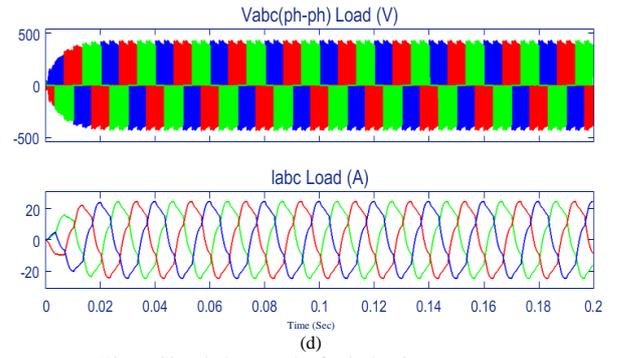

Fig. 7 Simulation results for induction generator.

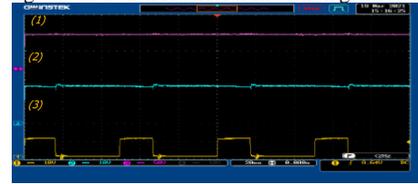

(a) (1) MOSFET gate pulse, (2) Input DC voltage, and (3) Output DC voltage.

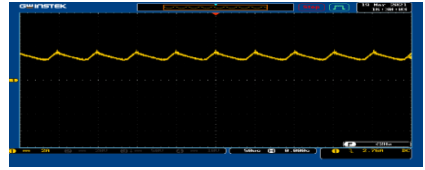

(b) Input current.

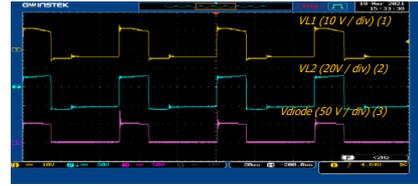

(c) (1) Voltage across the primary winding of the Y-source, (2) Voltage across the secondary winding of the Y-source, (3) Voltage across $D_1$.

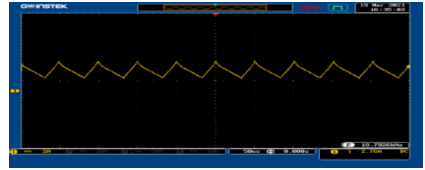

(d) Current through $L_r$.

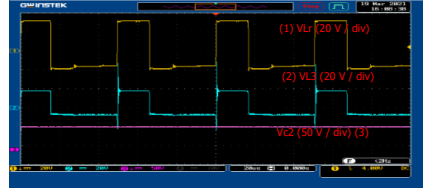

(e)
(1) $V_{Lr}$, (2) Voltage across the third winding of Y-source, (3) $V_{C2}$.
Fig. 8 Experimental results of the proposed converter for the resistive load.

## IV. CONCLUSION

One way to produce a flexible voltage gain, reduce the size of converter components, and achieve better short-circuit fault tolerance is by using Y-source converters. These features are critical to operational performance and safety in the propulsion systems of EVs and the integration of distributed energy storage, especially as it pertains to battery safety. A novel proposed topology with a low component count, simpler construct, and high voltage boost ratio was presented in this paper. Analytical and experimental results validate the performance of this converter for various types of loads to prove its high flexibility and resiliency with performance such as continuous input current, soft-switching during inductive loads, and decreasing startup inrush current.